# Embrace Opportunities and Face Challenges: Using ChatGPT in Undergraduate Students' Collaborative Interdisciplinary Learning

Gaoxia Zhu*
Assistant Professor
National Institute of Education (NIE)
Nanyang Technological University
E-mail: gaoxia.zhu@nie.edu.sg

Xiuyi Fan
Senior Lecturer
School of Computer Science and Engineering
Nanyang Technological University
E-mail: xyfan@ntu.edu.sg

Chenyu Hou
Ph.D. Student
Nanyang Technological University
E-mail: CHENYU004@e.ntu.edu.sg

Tianlong Zhong
Ph.D. Student
Nanyang Technological University
E-mail: tianlong001@e.ntu.edu.sg

Peter Seow
Research Scientist
National Institute of Education (NIE)
Nanyang Technological University
E-mail: peter.seow@nie.edu.sg

Annabel Chen Shen-Hsing
Professor
Division of Psychology
Nanyang Technological University
E-mail: annabelchen@ntu.edu.sg

Preman Rajalingam
Director
Centre for Teaching, Learning & Pedagogy

Nanyang Technological University
E-mail: prajalingam@ntu.edu.sg

Low Kin Yew
Associate Professor (Practice)
College of Business (Nanyang Business School)
Nanyang Technological University
E-mail: alowky@ntu.edu.sg

Tan Lay Poh
Associate Professor
School of Materials Science & Engineering
Nanyang Technological University
E-mail: lptan@ntu.edu.sg

Gaoxia Zhu and Xiuyi Fan contributed equally to this article.
**\* Corresponding author**
Learning Sciences and Assessment Academic Group, National Institute of Education (NIE),
Nanyang Technological University.
E-mail: gaoxia.zhu@nie.edu.sg

**Declaration**
**Competing interests:** The authors declare no potential conflict of interest in the work.

**Availability of data and materials:** Because of confidentiality agreements and ethical concerns, the data used in this study will not be made public. These data will be made available to other researchers on a case-by-case basis.
**Funding:** This study was supported by the NTU Edex Teaching and Learning Grants.
**Acknowledgments:** The authors are indebted to the students who participated in this study.

**Abstract**

ChatGPT, launched in November 2022, has gained widespread attention from students and educators globally, with an online report by Hu (2023) stating it as the fastest-growing consumer application in history. While discussions on the use of ChatGPT in higher education are abundant, empirical studies on its impact on collaborative interdisciplinary learning are rare. To investigate its potential, we conducted a quasi-experimental study with 130 undergraduate students (STEM and non-STEM) learning digital literacy with or without ChatGPT over two weeks. Weekly surveys were conducted on collaborative interdisciplinary problem-solving, and physical and cognitive engagement, and individuals reflected on their ChatGPT use. Analysis of survey responses showed significant main effects of topics on collaborative interdisciplinary problem-solving and physical and cognitive engagement, a marginal interaction effect between disciplinary backgrounds and ChatGPT conditions for cognitive engagement, and a significant interaction effect for physical engagement. Qualitative analysis of students' reflections generated eight positive themes about ChatGPT use, including efficiency, addressing knowledge gaps, and generating human-like responses, and eight negative themes, including generic responses, lack of innovation, and counterproductive to self-discipline and thinking. Our findings suggest that ChatGPT use needs to be optimized by considering the topics being taught and the disciplinary backgrounds of students rather than applying it uniformly. These findings have implications for both pedagogical research and practices.

**Introduction**

Released in November 2022, ChatGPT (Generative Pretrained Transformer, based on GPT-

3.5), a large language model pre-trained on massive text data using Reinforcement Learning from Human Feedback technique (Thorp, 2023), has received tremendous attention, interest, surprise, applause, and at the same time, concerns all around the world. In higher education, ChatGPT is shaking up the landscape. On the one hand, it brings new opportunities, such as triggering instructors to use ChatGPT to design innovative assessments and activities for teaching and learning purposes. On the other hand, the use of ChatGPT poses challenges, such as potential costs and efforts for educators to evaluate the relevance and accuracy of generated information and threats to student essays as evaluation assignments because ChatGPT-generated text takes only seconds to produce and may not trigger plagiarism detectors (Rudolph et al., 2023).

However, with a few exceptions (e.g., Ali, 2023; Shoufan, 2023; Yan, 2023), which we will discuss below, there is a lack of empirical research on the impact of ChatGPT in learning. Moreover, there is no ChatGPT study on collaborative interdisciplinary learning. Collaborative interdisciplinary learning involves using knowledge, methods, and insights from various disciplines to solve complex problems in challenging situations (Bybee, 2013; Ivanitskaya et al., 2002). In the first year of their university programs, undergraduate students may not have sufficient knowledge of their own or other disciplines, which hinders their ability to engage in interdisciplinary discussions and learning with other students. ChatGPT can be a valuable tool to complement students' lack of disciplinary knowledge because it is trained on a vast amount of data from different disciplines and has conversational capabilities and the ability to take on a specified persona or identity (Qadir, 2022).

In this work, we conducted an empirical research on the impact of ChatGPT with 130 undergraduate students enrolled in a digital literacy course in an Asian public university. The study lasted for two weeks, and the students worked on two learning topics, namely, Artificial

Intelligence (AI) and Blockchain. To examine the impact of ChatGPT on students' collaborative interdisciplinary learning and their perceptions, we collected student responses on class engagement and problem-solving skills using weekly surveys and online written self-reflections after each class. We analyzed the student data to identify positive and negative themes relating to their learning using ChatGPT. The benefits of this study are two-folded: (1) providing valuable insights into the potential benefits and challenges of using ChatGPT in collaborative interdisciplinary learning and (2) informing the development of effective teaching guidelines for students from different disciplines.

**Literature Review**

**ChatGPT in Education**

Chatbots are software programs that engage in real-time conversations with users (Clarizia et al., 2018). In education, they serve roles such as teaching agents, peer agents, teachable agents, and motivational agents (Kuhail et al., 2022). Chatbots can integrate multiple sources of information, answer students' questions immediately and motivate them (Okonkwo & Ade-Ibijola, 2021). They have been used in computing education, language learning, and mathematics education (Winkler et al., 2020; Jeon, 2021; Rodrigo et al., 2012). However, criticism such as lacking user-centered design, fictional conversations, and ethical issues are also there (Kuhail et al., 2022; Murtarelli et al., 2021).

ChatGPT is a powerful chatbot with human-like text processing abilities. Some are apprehensive about its use, while others are optimistic. In education, Kasneci et al. (2023) listed risks and challenges with using LLMs, including ethical and practical concerns. Tlili et al. (2023) raised concerns about ChatGPT's accuracy, fairness, and user privacy, while Qadir (2022) noted potential biases and ethical issues. Cheating and plagiarism are also concerns (Guo et al., 2023).

Rudolph et al. (2023) suggested that ChatGPT's limitations include understanding context, emotion, creativity, misinformation, response quality variation, and the danger of jailbreaking. Despite these concerns, there is a growing consensus that LLMs will become increasingly prevalent in daily life and learning (Looi & Wong, 2023; McMurtrie, 2022; Thorp, 2023; Tlili et al., 2023). Previous studies suggest that LLMs can have positive effects on providing automated assessments and adaptive feedback (Moore et al., 2022; Sailer et al., 2023; Zhu et al., 2020), stimulating curiosity (Abdelghani et al., 2022), increasing engagement (Tai & Chen, 2020), and supporting programming tasks and code explanations in computing education (Sarsa et al., 2022). ChatGPT has shown potential effectiveness in supporting language learning (Ali et al., 2023; Yan, 2023), medical education (Kung et al., 2023; Sallam, 2023), and engineering education (Qadir, 2022; Shoufan, 2023).

As ChatGPT is a relatively new technology, existing empirical research on its impacts on learning is limited. Rudolph et al. (2023) conducted a literature review on ChatGPT and higher education in January 2023 and found only two peer-reviewed articles and eight unreviewed preprints. The studies examined various applications of ChatGPT in learning, such as using it to write academic papers (Zhai, 2022) and having conversations with it (Qaidr, 2022). Other studies explored learners' perceptions of ChatGPT in language learning (Ali et al., 2023) and its potential to support L2 writing tasks (Yan, 2023). While some of the studies found that ChatGPT can be motivating and helpful to learning, others expressed concerns about its impact on academic integrity and educational equality (Yan, 2023; Shoufan, 2023). For example, some learners reported that ChatGPT improved their reading and writing skills but not their listening and speaking skills (Ali et al., 2023). Additionally, some students expressed concerns about the potential negative impacts of ChatGPT on academic integrity, personal life, and job prospects (Shoufan, 2023). Overall, while some studies suggest that ChatGPT may have benefits for learning, there are also concerns about its use. More research is needed to fully understand the impacts of ChatGPT on learning.

**Collaborative Interdisciplinary Learning**

Research has consistently shown that collaborative interdisciplinary learning offers a multitude of benefits to learners. For instance, it enables learners to personalize their organization of knowledge, develop critical thinking and metacognitive skills, and engage in meaningful collaboration (Alberta Education, 2015; Ivanitskaya et al., 2002; Ledford, 2015). Furthermore, this approach helps learners apply their knowledge and skills to real-world problems and enhances their ability to solve complex challenges effectively. Collaborative interdisciplinary learning is typically characterized by the integration of multiple disciplines, active collaboration towards shared goals, and the creation of innovative solutions across disciplinary boundaries. This approach also fosters the advancement of knowledge by identifying and solving new problems, explaining phenomena, and designing products (Klaassen, 2019; MacLeod & van der Veen, 2020).

The boundaries of different disciplines create a space where students can co-construct and develop new knowledge as well as contribute and integrate diverse ideas, but navigating this space is challenging because of disciplinary boundaries and conflicting epistemology (Akkerman & Bakker, 2011; MacLeod & van der Veen, 2020; Stentoft, 2017). Furthermore, forming collaborative interdisciplinary learning groups is challenging, as it is impossible to cover all the disciplinary knowledge needed for various learning problems and challenges. Difficulties include curriculum structures that focus more on single subjects, limited student representation from different disciplines, and students' limitations in terms of interdisciplinary knowledge, skills and methods (Authors, 2022; Ivanitskaya et al., 2002). To address these practical challenges, we explored whether ChatGPT could serve as additional group members to complement students' lack of disciplinary areas during collaborative interdisciplinary learning.

**The Current Study**

Despite the benefits of collaborative interdisciplinary learning in higher education, its implementation can be challenging. One potential solution to overcome this challenge is to leverage ChatGPT's ability to quickly access information from multiple disciplines. However, there is a limited amount of empirical research on the use of ChatGPT for education, and students' attitudes towards chatbots must also be taken into consideration (Okonkwo & Ade-Ibijol, 2021). Therefore, it is critical for researchers to conduct more studies to examine the appropriate use of ChatGPT, its potential benefits, challenges, and risks, as well as students' perceptions, particularly for collaborative interdisciplinary learning. Factors such as the learning content and students' disciplinary backgrounds and attitudes should be considered when incorporating ChatGPT in education. These investigations can inform the design of activities and materials for integrating ChatGPT into teaching and developing acceptable teaching guidelines (Qadir, 2022).

The purpose of our empirical study was to investigate the potential of ChatGPT as a peer agent to support collaborative interdisciplinary learning of two digital literacy topics, AI and Blockchain, among STEM and non-STEM undergraduate students. We designed a quasi-experiment involving 130 students from four tutorial groups and assessed their collaborative interdisciplinary problem-solving abilities, physical and cognitive engagement, and attitudes toward ChatGPT. Specifically, we aimed to answer the following research questions:

1. How do students' discipline backgrounds, ChatGPT conditions, and learning topics affect their collaborative interdisciplinary problem-solving abilities, physical and cognitive engagement?

2. Do STEM and non-STEM students differ in their attitudes toward using ChatGPT in collaborative interdisciplinary learning? What themes emerge in their reflections on ChatGPT use?

Our study aimed to provide valuable insights into the potential benefits and challenges of using ChatGPT in collaborative interdisciplinary learning and inform the development of effective teaching guidelines that have specifically considered students' disciplinary backgrounds.

**Methods**

**Participants**

This study was conducted in March 2023 in a digital literacy course offered to first or second-year undergraduate students at a public university in Asia. The participants were in four tutorial classes taught by the same instructor, with 48 students in tutorial 1 (i.e., T1), 48 in T2, 47 in T3, and 40 in T4 (183 students in total), respectively. The project was approved by the Institutional Review Board at the authors' institution, and students within these four tutorial classes were recruited. Among the 183 students, 130 (36 females, 54 males, 40 not indicated or did not fill in the survey, mean age=20.5) participated in this study. The participants enrolled in STEM programs (72 students from the College of Engineering and College of Science) and non-STEM programs (58 students from the College of Humanities and Social Sciences and College of Business). We then assigned students into groups of five to six members and ensured each group has a good representation of students from different schools to facilitate their interdisciplinary learning.

**Instructional Design**

The research team, comprising instructors, researchers, and postgraduate students with backgrounds in computer science, learning sciences, and psychology, developed two activities on AI and Blockchain topics. When designing the activities, we were guided by the characteristics of typical interdisciplinary problems synthesized from relevant literature (Kuo et al., 2019; Lam et al., 2014; MacLeod & van der Veen, 2020; Spelt et al., 2009). These considerations are outlined below.

Firstly, we aimed to create a real-world scenario for the students to make the task or problem authentic and relevant. For instance, in the AI module, we created a debating competition scenario in which interdisciplinary student teams prepared for a debate competition organized by the university debating society. The debate topic was "AI versus Internet: which one has a more profound impact on our society?"

Secondly, we encouraged collaborative contributions from various disciplines and made it explicit that students were expected to think from different disciplinary perspectives. For example, in the Blockchain module, students were tasked with studying successful Blockchain cases in different fields worldwide, learning from their experiences, and proposing commercial gaps that Blockchain could fill. We emphasized that the groups should not be constrained by a specific industry but can focus on the potential of ideas to make good use of their diverse disciplinary background asset.

Thirdly, we expected students to make authentic cognitive advancements, such as explaining a phenomenon, solving a problem, or designing a product that requires high-level cognitive efforts. In the AI module, we asked groups to collaboratively prepare for the debate by laying down arguments and counterarguments for each side.

Fourthly, the activities should not be easily solved by ChatGPT or other AI tools to ensure students' efforts in and control of their learning. For example, the debate topic of "AI versus Internet, which one has a more profound impact on our society?" does not have a straightforward answer. Therefore, students have to collaboratively work on the activities to deepen their understanding of the use of AI and the Internet in different fields, consider various factors and weigh different evidence and examples, and construct coherent and pervasive arguments while considering potential counterarguments from groups on the other side.

Finally, we provided scaffolding to support students in breaking down the tasks and using ChatGPT, as interdisciplinary learning tasks can be challenging for students to work on, and ChatGPT is a new tool for education. We broke down the tasks into smaller steps and provided scaffolding questions to guide students' learning and thinking. For instance, in the AI module, we segmented the debating task into four steps: identifying a framework, analyzing and evaluating applications, building arguments, and presenting argument points.

We used the same task in all four tutorial groups, either with or without ChatGPT support. We assigned T1, T2, T3, and T4 to either ChatGPT or non-ChatGPT conditions based on two criteria: 1) including a control group without using ChatGPT, and 2) giving each tutorial at least one opportunity to participate in the ChatGPT condition during the two-week experiment. Table 1 shows the arrangement of conditions for each tutorial. It is important to note that the ChatGPT condition included two options: normal ChatGPT, which allowed students to use it freely without constraints or instructions, and ChatGPT Persona, which involved defining a persona (e.g., Jack, a 23-year-old Singaporean first-year undergraduate from the business school) to provide additional disciplinary knowledge to the group and involve them in discussions and problem-solving.

**Table 1.**
*Condition arrangements for students in different tutorials in the two-week quasi-experiment*

| Tutorials | AI week | Blockchain week |
|---|---|---|
| T1 | ChatGPT Persona | ChatGPT Persona |
| T2 | Non-ChatGPT | ChatGPT |
| T3 | ChatGPT | Non-ChatGPT |
| T4 | ChatGPT | ChatGPT |

A flipped classroom learning approach was implemented for the course delivery. Prior to each class, students were instructed to watch an hour-long pre-recorded video on the week's topic (i.e., AI and Blockchain) and complete a short self-assessment test. During class, in each tutorial, small groups with STEM and Non-STEM students were provided with two hours to engage in collaborative interdisciplinary learning activities, with or without the use of ChatGPT based on their assigned treatment conditions. The students participated in face-to-face activities with the aid of Miro, a visual collaboration platform (https://miro.com, see Figure 1). Following each class, students were required to write individual reflections.

**Figure 1.**
*Content Board of Introduction to Artificial Intelligence*

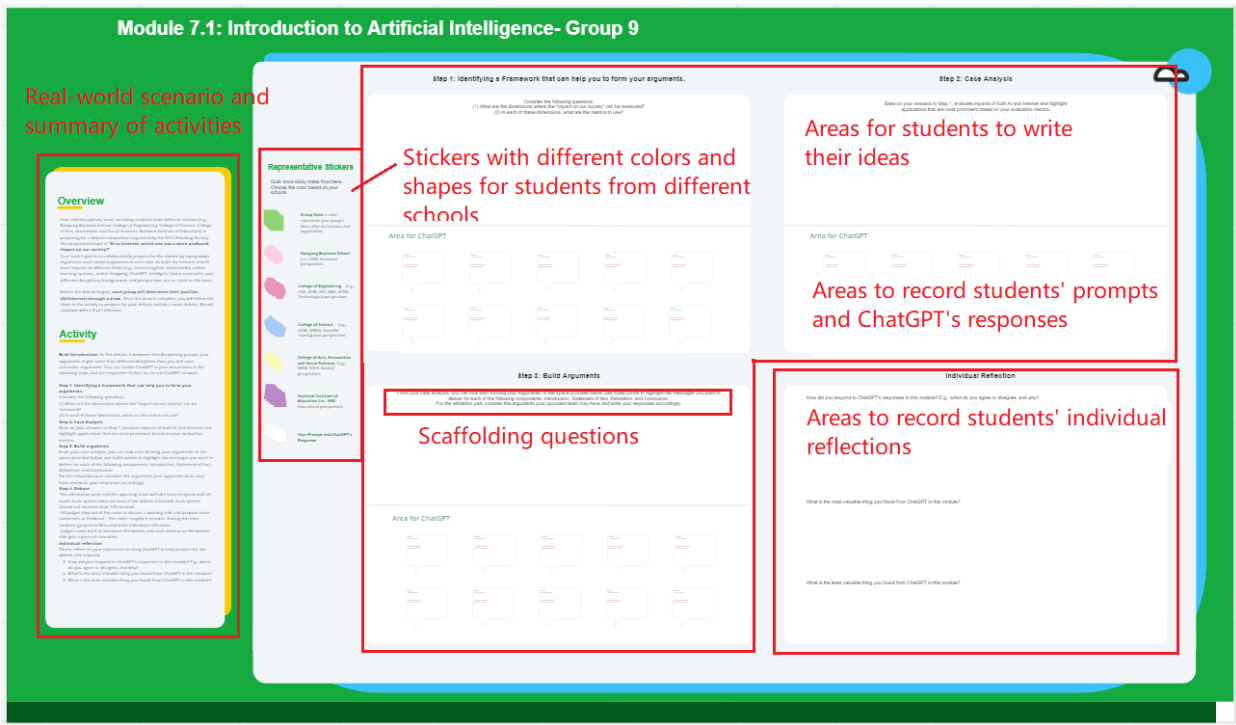

**Learning Environment**

In Miro, we used navigation and content boards for each tutorial and group, respectively,

as shown in Figure 1. We included stickers of various shapes and colors to track the contributions of students from different schools and ensure inclusivity for color-blind students as well as provided group note stickers to encourage collaboration. We also provided a breakdown of the steps and detailed questions for each step to facilitate the learning process. To remind students to record their interactions with ChatGPT, we separated their response areas and ChatGPT areas, and placed stickers for ChatGPT's input on the template.

**Data Collection**

This study included three data sources: (1) students' age, gender and disciplinary background information collected at the beginning of the semester, (2) weekly surveys on students' collaborative interdisciplinary problem-solving, and physical and cognitive engagement and (3) 167 pieces of written reflections on the use of ChatGPT by individuals. Next, we would elaborate on how the surveys and individual reflections were collected.

*Collaborative interdisciplinary problem-solving survey*

In each week's tutorial, the participants were asked to complete surveys on their collaborative interdisciplinary problem-solving and engagement using google forms after the group activities and students' individual reflections. Collaborative interdisciplinary problem-solving was measured by three questions adopted from the Integration subscale in the Team Collaboration Questionnaire (Cole et al., 2018) on a five-point Likert scale (1 represents strongly disagree and 5 strongly agree). The questions include "I investigated an issue with my team to find an acceptable solution," "I integrated my ideas with my team to come up with a joint solution," and "The solution my team and I found to the problem satisfies our needs." The Cronbach's alpha coefficient for this subscale is .91 (Cole et al., 2018).

*Physical and cognitive engagement survey*

Engagement (i.e., physical and cognitive engagement) was measured by four questions adapted from Student Engagement (Gerald et al., 2015). This questionnaire measures students' physical and cognitive engagement in class on a five-point Likert scale (1 denotes strongly disagree and 5 strongly agree). Questions to measure physical engagement include "I devoted a lot of energy toward the group activities today" and "I tried my hardest to perform well in the group activities today"; cognitive engagement is measured by "When working on the group activities, I paid a lot of attention" and "when working on the group activities, I concentrated on the work." The Cronbach's alpha coefficient is .93 for physical engagement and .96 for cognitive engagement (Gerald et al., 2015).

*Individual reflection on the use of ChatGPT*

To gain insight into students' experience with ChatGPT, we collected their reflections immediately after completing learning tasks that involved ChatGPT. In each tutorial, we asked individuals to respond to predetermined questions in a designated reflection area on the Miro board, as shown in Figure 1. To ensure the learning processes and components were as similar as possible for students in the non-ChatGPT condition, we also asked them to reflect on two questions in Miro after their learning activities. Table 2 displays the questions that guided students' reflections in both the ChatGPT and non-ChatGPT conditions. However, it should be noted that this study focuses exclusively on reflections from the ChatGPT condition to gain insights into students' opinions and reflected themes.

**Table 2.**
*Individual reflection questions for the ChatGPT and non-ChatGPT conditions*

| Individual reflection questions for the ChatGPT condition | Individual reflection questions for the non-ChatGPT condition |
| --- | --- |
| 1. How did you respond to ChatGPT's responses in this module? E.g., when do you agree or disagree, and why?<br>2. What is the most valuable thing you found from ChatGPT in this module? | 1. List one thing that your group has done well in relation to all other groups.<br>2. List one thing you will improve in your next debate exercise. |

| 3. What is the least valuable thing you found from ChatGPT in this module? |
|---|

**Data Analysis**

**ANOVA**

We employed ANOVA and factorial ANOVA to answer RQ1 and investigate the main and interaction effects of students' disciplinary backgrounds, ChatGPT conditions, and learning topics on collaborative interdisciplinary problem-solving, physical engagement, and cognitive engagement. Due to sample size and statistical power considerations, we categorized students' disciplinary backgrounds into STEM (engineering and science) and non-STEM programs (humanities and social sciences, and business). We also classified conditions as ChatGPT (ChatGPT and ChatGPT Persona) and non-ChatGPT. First, we examined the main effect of topics (AI & Blockchain), ChatGPT conditions, and disciplinary backgrounds on students' collaborative interdisciplinary problem-solving, physical engagement, and cognitive engagement. We then conducted a factorial ANOVA to investigate the interaction effects of students' learning topics, ChatGPT conditions, and disciplinary backgrounds on the dependent variables.

**Sentiment analysis and qualitative analysis**

To answer RQ2, which investigates potential differences in STEM and non-STEM students' attitudes towards using ChatGPT during their collaborative interdisciplinary learning and the themes emerging from their reflections, we performed sentiment analysis using the DistilBERT model (Sanh et al., 2019). Each reflection was labeled as either positive or negative by the model. The third author reviewed the sentiment analysis results and highlighted the ones different from her interpretation, resulting in 17 out of 167 reflections being highlighted. The first author then reviewed the highlighted reflections and discussed them with the third author until they reached a consensus. The agreed sentiment analysis results were subjected to a t-test to examine if STEM and non-STEM students had different attitudes toward using ChatGPT.

To examine the themes that emerged from students' reflections, the first author coded the reflections on using ChatGPT during collaborative interdisciplinary learning. First, the author separated the student reflections from T1 into relatively positive and negative ones. Next, the author synthesized the coding and identified common themes through an inductive reasoning approach, a "bottom-up" analytical strategy (Thorne, 2000). Inductive reasoning allows researchers to use the data to generate ideas by interpreting and structuring the meanings derived from data (Thorne, 2000). Specifically, the first author first labeled the theme "Efficiency" which was mostly mentioned by students and then read through the remaining data to see if they fit the existing identified themes. If yes, the relevant quotes would be labeled with the theme; otherwise, new themes will be added until all the data has been assigned a theme. Using the same process, the author analyzed data from other tutorials to examine whether new themes needed to be added until all reflections had been analyzed. The third author then read through all the reflections, checked the reflections against the themes, and confirmed the themes identified by the first author was appropriate and comprehensive.

**Results**

**RQ1: Main and interaction effects of students' disciplinary backgrounds, ChatGPT conditions and learning topics**

Table 3 displays the descriptive statistics for collaborative interdisciplinary problem-solving, physical engagement, and cognitive engagement of STEM and non-STEM students over two weeks of classes in ChatGPT and non-ChatGPT conditions. STEM students had higher average scores in collaborative interdisciplinary problem-solving, physical engagement, and cognitive engagement under the ChatGPT condition than non-STEM students. The only exception is that in the Blockchain module, non-STEM students had higher cognitive engagement (mean score of 3.91) than STEM students (mean score of 3.83). Non-STEM students had relatively higher average scores on the dependent variables in the non-ChatGPT condition.

**Table 3.**

*Descriptive statistics of physical engagement, cognitive engagement and collaborative interdisciplinary problem-solving*

| Topics | | ChatGPT | | Non-ChatGPT | |
| --- | --- | --- | --- | --- | --- |
| | | STEM | Non-STEM | STEM | Non-STEM |
| AI & Internet | Number of Participants | 69 | 50 | 27 | 21 |
| | Number of Responses | 45 | 36 | 10 | 9 |
| | Collaborative Interdisciplinary Learning M(SD) | 4.24 (.59) | 4.15 (.77) | 4.06 (.51) | 4.22 (1.16) |
| | Physical Engagement M(SD) | 4.21 (.60) | 4.02 (.79) | 4.20 (.63) | 4.55 (.72) |
| | Cognitive Engagement M(SD) | 4.177 (.06) | 3.95 (.84) | 4.15 (.33) | 4.55 (.72) |
| Blockchain | Number of Responses | 46 | 35 | 17 | 12 |
| | Collaborative Interdisciplinary Learning M(SD) | 3.84 (.69) | 3.79 (.63) | 3.82 (.71) | 4.13 (.59) |
| | Physical Engagement M(SD) | 3.83 (.82) | 3.74 (.61) | 3.52 (.81) | 4.08 (.66) |
| | Cognitive Engagement M(SD) | 3.83 (.85) | 3.91 (.69) | 3.76 (.73) | 4.16 (.57) |

As shown in Table 4, the factorial ANOVA results indicate significant main effects of topics on collaborative interdisciplinary problem-solving ($F(1, 202) = 13.19, p < .01, \eta^2 = .05$), cognitive engagement ($F(1, 202) = 5.91, p < .05, \eta^2 = .02$), and physical engagement ($F(1, 202) = 15.15, p < .001, \eta^2 = .06$). A post hoc Tukey test shows students' collaborative interdisciplinary learning, physical engagement, and cognitive engagement were consistently higher for the AI & Internet topic. Different ChatGPT conditions did not have a main effect on the dependent variables.

**Table 4.**
*ANOVA results: Main and interaction effects of students' disciplinary backgrounds, ChatGPT conditions, and learning topics*

| | Main Effect | | | Interaction Effect | | | |
| --- | --- | --- | --- | --- | --- | --- | --- |
| | Disciplinary Background $F(1, 202)$ | Condition $F(1, 202)$ | Topic $F(1, 202)$ | Condition × Disciplinary Background $F(1, 202)$ | Topic × Disciplinary Background $F(1, 202)$ | Topic × Condition $F(1, 202)$ | Topic × Disciplinary Background × Conditions $F(1, 202)$ |
| Collaborative Interdisciplinary Problem-solving | .00 | .20 | 12.40*** | 1.83 | .10 | .85 | .06 |
| Physical Engagement | .00 | .45 | 15.15**** | 6.29** | .31 | 1.05 | .05 |
| Cognitive Engagement | .12 | 1.46 | 5.91** | 3.37* | 1.24 | .52 | .36 |

*Note.* Topics: AI & Internet versus Blockchain; Conditions: ChatGPT versus non-ChatGPT; Disciplinary Backgrounds: STEM versus non-STEM.
*$p < .1$; **$p < .05$; ***$p < .01$; ****$p < .001$

Regarding interaction effects, we found a marginal interaction effect between students' disciplinary background and ChatGPT condition for cognitive engagement ($F(1, 202) = 3.37, p = 0.06, \eta^2 = .01$), and a significant interaction effect for physical engagement ($F(1, 202) = 6.29, p < .05, \eta^2 = .03$). As plotted in Figure 2, the post hoc comparisons using the t-test with Bonferroni correction show that non-STEM students were more physically and cognitively engaged during collaboratively learning in the non-ChatGPT condition. No significant difference in the level of engagement was found for STEM students between ChatGPT and non-ChatGPT conditions.

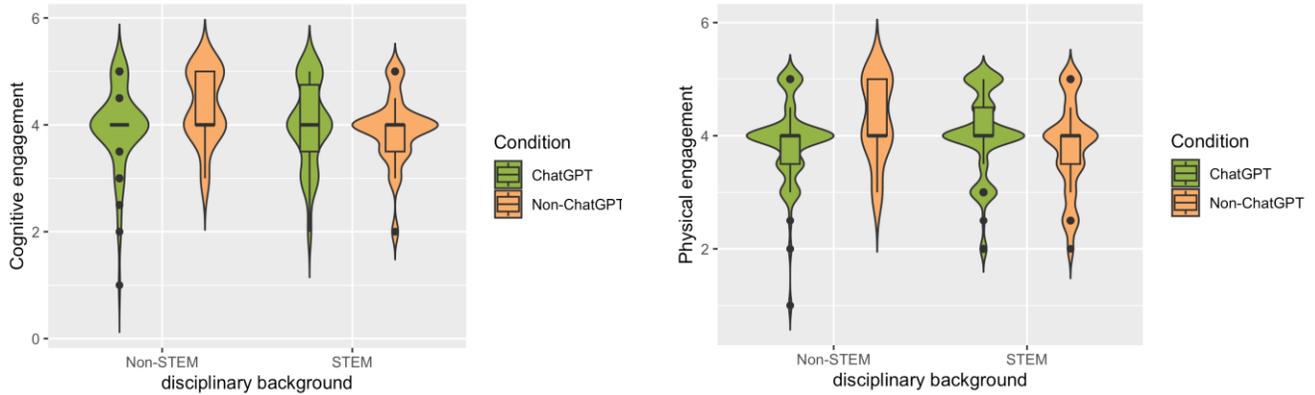

**Figure 2.**
*Interaction effects between disciplinary backgrounds and ChatGPT conditions on physical engagement and cognitive engagement*

**RQ2: STEM and non-STEM Students' opinions on the use of ChatGPT and the themes that emerged in their reflections**

In our sentiment analysis, we assigned a value of "1" to all positive posts and "-1" to negative posts, and conducted a t-test to compare the sentiment scores of STEM and non-STEM students. The results showed that there was no significant difference in the opinions of STEM and non-STEM students on the use of ChatGPT in collaborative interdisciplinary learning ($t(137.2) = .92$, $p = .35$). However, we wanted to gain a deeper understanding of their perspectives, so we analyzed all individual reflections from STEM and non-STEM students. Table 5 presents the eight positive and eight negative themes that emerged from our analysis. Each theme is described in detail below.

**Table 5.**
*Positive and negative themes generated from students' individual reflections*

| Positive themes | Negative themes |
| --- | --- |
| P1: Efficiency | N1: Inaccuracy, fact-checking and verification required |
| P2: Accurate and insightful, unbiased, | N2: Misleading and biased decision-making |
| reasonable and justified, structured and comprehensive and targeted responses | |
| P3: Providing different perspectives and supporting brainstorming and idea generation | N3: Generic but not contextualized, brief but not profound; therefore, additional probing or research is required |
| P4: Addressing knowledge gaps and complementing thinking | N4: Depending on prompts and having usage context limits |
| P5: Inspiration and refinement through further research and probing | N5: Repetition, wordiness and no innovative or original ideas |
| P6: Not judgmental and can generate human-like responses | N6: Harming self-discipline and thinking |
| P7: Safeguards for its misuse | N7: Human labor replacement |
| P8: Providing language support | N8: Technical issue |

*Positive themes*

**P1: Efficiency**. In students' reflections, a frequently mentioned theme is the efficiency of ChatGPT. The participants were surprised at how quickly and efficiently ChatGPT responded to their input/prompts and consolidated information from different sources. As a result, it saves students' time. For instance, in the AI module, a STEM student was surprised by ChatGPT's speed by exclaiming, "ChatGPT is able to give us our desired information at O(1) speed … Insane time complexity!" Similarly, in the same module, a non-STEM student reflected that "ChatGPT … reduces the time needed for humans to analyze the different sources of data and compile it together to form a concrete argument."

**P2: Accurate and insightful, unbiased, reasonable and justified, structured and comprehensive and targeted responses**. Students found ChatGPT's responses accurate, insightful, logical, reasonable, unbiased, structured, comprehensive, and customized. One STEM student praised ChatGPT's accuracy in the AI module, while a non-STEM student appreciated its ability to offer "a lot of insightful responses from a different perspective." Students noted ChatGPT's unbiased and neutral approach, with one stating, "ChatGPT took a neutral side." Students also appreciated ChatGPT's justification, comprehensiveness, and structure. One STEM student in the Blockchain module noted that ChatGPT "helps to put structure to scattered ideas and

thoughts," while a non-STEM student appreciated the bot's comprehensive responses with reasons to justify them. Some students believed ChatGPT customized its answers to their questions.

**P3: Providing different perspectives and supporting brainstorming and idea generation**. Participants commonly mentioned ChatGPT's ability to provide different perspectives, which they found helpful. One STEM student in the AI module noted, "ChatGPT provides a very broad answer that gives the users a good idea of the different information surrounding the questions they asked." In the Blockchain module, a STEM student found it useful as an idea generator, saying, "ChatGPT offers a broad range of content that even I wouldn't think of." ChatGPT's affordances make it appropriate for supporting users to brainstorm ideas and start their work efficiently and comprehensively. For instance, a student reflected in the AI module that "with the key points/factors that ChatGPT generates, it was easy to get started on the research."

**P4: Addressing knowledge gaps and complementing thinking**. Some students found ChatGPT to be useful for addressing knowledge gaps and complementing thinking. According to a STEM student in the AI module, "It can address gaps in technical knowledge, allowing for those who don't have that knowledge to apply it similarly." Another STEM student in the Blockchain module reflected, "ChatGPT uncovered information that I usually would not think of; it helps me to see other perspectives/applications." Similarly, some non-STEM students found that ChatGPT "found other applications of blockchain technology other than crypto and finance" and "can help to complement our ideas and thinking."

**P5: Inspiration and refinement through further research and probing**. Some students found that ChatGPT inspires them and allows them to build on its responses and do further research to improve the responses. One non-STEM student in the AI module responded, "The responses are very insightful (when I have really no ideas), leading me to think more and build more opinions on them." Similarly, a STEM student wrote, "I feel like ChatGPT inspired me to think broadly, as if I was talking to another person. ChatGPT allowed me to dive into another thought process, giving me a more complete look into the issue." ChatGPT's ability to retain conversation history allows users to "ask it things that you have previously mentioned." However, some students indicate the need for further research rather than total dependence on ChatGPT.

**P6: Not judgemental and can generate human-like responses**. Participants like ChatGPT because it can generate human-like responses and serve as a supporting and non-judgmental teammate. One STEM student indicated, "ChatGPT inspired me to think broadly, as if I was talking to another person." Another STEM student added, "ChatGPT is very informative and can construct logical explanations eerily similar to how a human can process thought and decision making." A non-STEM student suggested what they liked most about ChatGPT was its "ability to generate human-like responses." Another non-STEM student discussed that being "committed, knowledgeable, caring, willing to participate, does not judge you no matter how stupid your question is makes ChatGPT a good teammate."

**P7: Safeguarding for its misuse**. Although rarely mentioned by the participants, another distinct theme that students like about ChatGPT is its affordance of safeguarding its misuse. Specifically, in the AI module, a STEM student wrote, "ChatGPT has safeguards for its misuse or its use for harmful/criminal activities."

**P8: Providing language support**. Although the two activities do not emphasize spelling or grammar issues, after the activities, a few students reflected that ChatGPT supported their language use ("It also useful when it comes to language matters") or helped them check grammar ("We will agree to use ChatGPT during the research stage and use it for grammar checking").

*Negative themes*

**N1: Inaccuracy, fact-checking and verification are required**. One major concern students expressed regarding ChatGPT is its possibility of misinterpreting inputs and providing unreliable or false answers. In the AI module, a non-STEM student expressed concern: "ChatGPT may sometimes be unable to understand our question and hence not generate an accurate answer." Similarly, in the Blockchain module, students reflected that "credibility needs to be verified" and "not everything they explain is always correct." Moreover, some students acknowledged that ChatGPT's responses might not be up-to-date and that it could be challenging to fact-check them. As a result, students emphasized the need for fact-checking to ensure the accuracy of ChatGPT's responses and to prevent misinformation. However, they also recognized that this could be difficult, given ChatGPT's extensive knowledge and the need to understand the topic to fact-check its responses.

**N2: Misleading or biased decision-making**. Students expressed concerns about ChatGPT's potential for providing inaccurate or unreliable responses, leading to misunderstanding information and biased decision-making. In the AI module, one student noted, "ChatGPT may provide an insight more inclined to whoever is feeding ChatGPT the information." Similarly, in the Blockchain module, a student mentioned that "ChatGPT is limited to information before 2021 and only has sparse knowledge on real-world events after that year." Another student highlighted that ChatGPT could return information that supports the user's perspective, potentially ignoring other sides of the argument. Additionally, some users may not be able to detect mistakes made by ChatGPT due to their limited understanding of the topic. As a result, students emphasized the importance of fact-checking and verifying ChatGPT's responses to prevent misinformation and bias.

**N3: Generic but not contextualized, brief but not profound; therefore, additional probing or research is required**. Students have concerns about the limitations of ChatGPT, with

some citing its tendency to provide generic, vague, and superficial responses that lack context, specificity, and depth. One student noted that "ChatGPT doesn't do well in terms of giving exact case studies, but only gives a generic answer to questions." Some students also felt that ChatGPT's responses lacked depth, with comments such as "The responses are usually very brief and not very in-depth" and "the lack of insights in the information." Therefore, students suggested that additional research or using tools like "google scholar, scientific websites..." could help obtain more in-depth analyses.

**N4: Depending on prompts and having usage context limits**. Students have raised concerns about ChatGPT's limitations, particularly in generating comprehensive responses. Two non-STEM students in the AI module noted, "ChatGPT does not provide information fully at one go, several prompts are needed to force AI to think" and "an answer may not really answer the question if not prompted correctly." Participants generally agree that learners need to adjust their prompts to get comprehensive responses from ChatGPT. Some students also pointed out that ChatGPT has limitations in evaluating complex concepts, tasks requiring personal opinions, and topics requiring nuanced perspectives. For example, a STEM student in the AI module commented, "ChatGPT is not very useful to draw insights on concepts that are tough to evaluate, lacking a certain form of intelligence when it comes to personal opinion matters."

**N5: Repetition, wordiness and no innovative or original ideas**. Some students are concerned about the repetitiveness and wordiness of ChatGPT's responses. They suggest refining ChatGPT's responses by filtering out unneeded information and avoiding excessive repetition. For instance, students recommended asking ChatGPT to "narrow down/cut out unneeded information." Students also expressed frustration with ChatGPT's limitations in generating new ideas and original content, as one student noted that ChatGPT struggles to "come up with new ideas, in terms of innovation and original ideas."

**N6: Impairing self-discipline and thinking**. The participants raised concerns about how ChatGPT negatively affected their critical thinking and effort to generate ideas. In the AI module, a STEM student noted, "ChatGPT gives concise answers which take away the need for critical thinking." Similarly, in the Blockchain module, a STEM student mentioned that "it stops me from thinking," and another student reflected, "ChatGPT sounded so confident so I just believed in him and took his responses and based my thoughts on it, showing the influence and impact AI has on us." The students recognized that relying too much on ChatGPT could make them lazy to think for themselves, with one STEM student noting, "repeatedly using ChatGPT can be a bad habit as it makes us lazy to think (by) ourselves." Others expressed similar concerns, stating that "ChatGPT makes me tempted not to expend effort to generate ideas myself" and "My brain didn't think much because AI is too professional." However, the participants also acknowledged the need for self-discipline, with one saying they had been "too dependent on it," and another acknowledging "there is the urge to be reliant on ChatGPT."

**N7: Human labor replacement**. A few participants were worried that ChatGPT was so capable of providing wholesome and structured responses that it might replace their jobs. In the AI module in the ChatGPT persona condition in which student groups could define persona to support and complement their debating teams, two students reflected that "What extra value can we bring? I might as well make 6 different personas and ask ChatGPT to debate. I think it will do a better job" and "ChatGPT may replace jobs. This is a concern that can be readily addressed."

**N8: Technical issue**. A few participants encountered technical or stability issues with ChatGPT (e.g., network errors or unknown reasons) in the Blockchain module. They included these issues in their reflections and indicated they used Google instead because "For me, ChatGPT didn't really work due to network errors as they told me. Only managed to answer 1 question before it failed again" and "ChatGPT not working due to some unknown reason, still faster to google."

**Discussions**

The impact of ChatGPT on students' in-class collaborative interdisciplinary problem-solving and engagement was examined. While self-reported engagement and collaborative interdisciplinary learning did not differ between ChatGPT and non-ChatGPT conditions, quantitative analysis revealed higher levels of collaborative interdisciplinary problem-solving, physical engagement, and cognitive engagement in the AI module compared to the Blockchain module. This result is consistent with prior research on the importance of instructional design in computer-supported collaborative learning (Luckin & Cukurova, 2019). In detail, the AI module involved formulating arguments to defend a side in a debate, whereas the Blockchain module required conducting a case analysis, which mainly involved factual information that students could adopt from websites or ChatGPT responses. Our result aligns with a meta-analysis showing that computer-supported collaborative learning can harm case-based learning (Jeong et al., 2019), where learners might overly rely on website information and compromise interactions with peers (Kemp et al., 2019). Hence, instructors need to consider preserving students' agency and critical thinking while maximizing the advantages of ChatGPT (e.g., efficiency, comprehensive, and targeted responses) when designing learning activities.

We found that non-STEM students tended to be more physically and cognitively engaged in the learning activities without help from ChatGPT. In contrast, STEM students showed no significant differences in collaborative interdisciplinary problem-solving and engagement between ChatGPT and non-ChatGPT conditions. Previously, Jeong et al. (2019) found that the disciplinary backgrounds of students moderate the effect of technology on learning outcomes. Thongsri et al.

(2019) found that non-STEM students, especially those with low computer self-efficacy, have a lower behavioral intention to use new learning technologies. We conjecture that non-STEM students might spend extra time getting familiar with ChatGPT functions and were left out during group discussions. STEM students, on the other hand, have high digital literacy (Thongsri et al., 2019) and are more comfortable integrating new technology, such as ChatGPT, into their learning process. They naturally treat ChatGPT as a new form of seeking information and can integrate the ChatGPT responses into their problem-solving process as they do with traditional web searches. However, more research is needed to understand how students from different disciplines respond to ChatGPT during learning activities and how the unevenness of digital literacy influences collaborative problem-solving.

The sentiment analysis results showed no significant difference in the perceptions towards using ChatGPT between non-STEM and STEM students. However, it is important to consider how we framed the reflection questions, as shown in Table 2. By asking participants to reflect on the most valuable and least valuable aspects of using ChatGPT in the module, we may have inadvertently biased the responses towards positive and negative perceptions, regardless of the disciplinary background. In contrast, Shoufan (2023) asked undergraduate students an open-ended question, "What do you think of ChatGPT? Think deeply and write down whatever comes into your mind!" and found that almost 67% (56/171) of the comments were positive. To gain a deeper understanding of students' perceptions towards using ChatGPT in learning, future studies may consider using more open-ended and less biased methods, such as semi-structured interviews.

We qualitatively analyzed students' reflections on their experience using ChatGPT and identified several positive themes. First, ChatGPT provided students with different perspectives, supported brainstorming and idea generation, and addressed knowledge gaps. Second, ChatGPT complemented thinking and inspired students to refine their ideas through further research and probing. Third, students appreciated ChatGPT's non-judgmental nature and ability to generate human-like responses. These findings are consistent with McMurtrie's (2023) suggestion that AI tools may "help ignite the imaginative process." Shoufan (2023) also identified similar positive themes, such as ChatGPT being "helpful for learning" and "helpful for work" and generating human-like quality of conversations.

Based on students' reflections, some negative impacts of ChatGPT on learning were identified, including the potential for misleading or biased decision-making, repetitive or wordy responses, and a lack of innovative or original ideas. Tlili et al. (2023) found that learners reported similar concerns about ChatGPT's responses being potentially misleading or even eroding their creative and critical thinking skills. Qadir (2022) also noted that ChatGPT can generate misinformation due to limitations in LLM models, training data quality, and prompt inputs. OpenAI itself warns that ChatGPT can give seemingly confident but potentially unreliable answers to complex questions (OpenAI, 2022), and CEO Sam Altman has cautioned against overreliance on ChatGPT, stating that it is not yet reliable enough for anything important (Altman, 2022). As Rudolph et al. (2023) suggested, ChatGPT may be less effective at handling content that requires critical thinking and analysis and should be seen as a tool to complement and enhance learning rather than replace teachers' and students' roles. Students in this study also highlighted the importance of developing literacy to evaluate AI-generated content. They emphasized that AI cannot replace the learning process and self-discipline is necessary for critical thinking.

The prevailing notion is that students will plagiarize content generated by ChatGPT and integrate it into their academic assignments. However, the findings from the study indicate that students are aware of the limitations of ChatGPT, as evidenced by the negative feedback they provide, without the need for explicit instruction from educators. The findings are positive indicators from an instructional standpoint, with implications for more effective pedagogy. Rather than prohibiting the use of ChatGPT, educators can initiate a dialogue with students about the challenges associated with its use. Allowing students to utilize ChatGPT in the classroom and facilitating discussions about its potential benefits and drawbacks could be a constructive approach.

On the other hand, only a small number of students reflected on ethical considerations associated with using ChatGPT in their learning. OpenAI's efforts to avoid offensive outputs may have led students to overlook ethical considerations. OpenAI uses reinforcement learning, human feedback, recursive reward modeling, and moderation models to ensure the safe use of ChatGPT (Ouyang et al., 2022; Markov et al., 2022). However, ethical considerations should still be addressed in the use of AI technology, and future research should explore how to better incorporate these considerations into the learning experience.

ChatGPT and other AI models have the potential to revolutionize education, but it is important to carefully consider their impact on students' self-discipline and critical thinking, as well as potential risks such as repetitive and conventional ideas. Additionally, factors such as learning content, instructional design, and students' differences should be considered when implementing ChatGPT in education. As such, there is a need for collective efforts to research ChatGPT and other forthcoming AI tools at philosophical, epistemic, and empirical levels to ensure they do more good than harm in the rapidly changing era.

**Implications**

ChatGPT has emerged as a significant technological advancement in education. While students embrace ChatGPT, they are also aware of its potential drawbacks. This study has practical and epistemological implications for future research and practice in higher education.

Epistemologically, educators need to rethink critical questions, such as what competencies students should develop in the era of powerful AI, how to consider knowledge generated by ChatGPT, and how to anticipate potential ethical issues. At a practical level, researchers need to explore activities and assessments that can help students develop and evaluate required competencies and balance the relationships between students, instructors, and AI in classrooms. These questions and issues need to be addressed to ensure that ChatGPT and other AI tools are used effectively in education.

**Limitations and future directions**

The current study has several limitations that should be addressed in future research. First, although the sample size of the study is decent, the participants were only first-year undergraduate students who participated in two relatively similar activities on AI and Blockchain in a digital literacy course at a single Asian university. To generalize the findings, future research should survey more senior students with different cultural backgrounds and students participating in different learning activities in various courses over a longer period of time. Additionally, researchers should consider the nuances between using ChatGPT in general and using ChatGPT persona, which allowed students to define their "teammate." Second, the study focused on students' perceived benefits and drawbacks of ChatGPT after hands-on experience in class, but did not analyze their interaction data with ChatGPT. Therefore, future studies should investigate how ChatGPT may influence learning positively or negatively by analyzing students' interaction data with ChatGPT. Moreover, researchers should design learning activities that leverage ChatGPT and assess students' learning in its context. Finally, although the study followed the five characteristics of typical interdisciplinary problems for both the AI and Blockchain modules, there might still be some nuanced differences between the two tasks. Future research should consider these subtle differences between task design and topics when applying ChatGPT or other AI tools.

**Conclusion**

ChatGPT has garnered significant attention in higher education due to its potential to provide personalized and interdisciplinary learning opportunities. However, despite researchers' hypotheses about ChatGPT's impact, there is a lack of empirical research on its application in controlled classroom settings, particularly in interdisciplinary contexts. To address this research gap, we conducted a two-week quasi-experimental study with 130 undergraduate students, examining their collaborative interdisciplinary problem-solving, engagement, and perceptions of ChatGPT in a digital literacy course at an Asian public university. Our findings suggest that to optimize the use of ChatGPT, educators need to consider both the disciplinary backgrounds of students and the topics being taught (or how the activity is designed), as students with different backgrounds react differently to ChatGPT across different topics. Furthermore, our analysis of students' reflections identified positive and negative themes regarding using ChatGPT, providing valuable insights for educators seeking to integrate ChatGPT into their teaching. Overall, this work contributes to the limited empirical research on ChatGPT's use in higher education and provides practical insights into optimizing its application.